\documentclass[runningheads]{llncs}
\usepackage[T1]{fontenc}
\usepackage{graphicx}
\usepackage{hyperref}
\usepackage{color}

\let\svthefootnote\thefootnote
\newcommand\freefootnote[1]{%
  \let\thefootnote\relax%
  \footnotetext{#1}%
  \let\thefootnote\svthefootnote%
}

\begin{document}
\title{An automated pipeline for quantitative T2* fetal body MRI  and segmentation at low field}

\titlerunning{Low Field Quantitative T2* Fetal Body MRI}

\author{Kelly Payette \inst{1,2,*} \and
Alena Uus\inst{1,2} \and
Jordina Aviles Verdera\inst{1,2} \and
Carla Avena Zampieri\inst{1,2} \and
Megan Hall\inst{1,3} \and
Lisa Story\inst{1,2} \and
Maria Deprez\inst{1,2} \and
Mary A. Rutherford\inst{1} \and
Joseph V. Hajnal\inst{1,2} \and
Sebastien Ourselin\inst{1,2} \and
Raphael Tomi-Tricot\inst{2,4} \and
Jana Hutter\inst{1,2}}

\authorrunning{Payette et al.}

\institute{Centre for the Developing Brain, School of Biomedical Engineering \& Imaging Sciences, King's College London, London, UK
 \and
 Department of Biomedical Engineering, School of Biomedical Engineering \& Imaging Sciences, King's College London, London, UK \and
Department of Women \& Children's Health, King's College London, London, UK:
MR Research Collaborations, Siemens Healthcare Limited, Camberley, UK \\$*$Corresponding author: \email{kelly.m.payette@kcl.ac.uk}
}

\maketitle
\freefootnote{K. Payette, A. Uus - Authors contributed equally}

\begin{abstract}
Fetal Magnetic Resonance Imaging at low field strengths is emerging as an exciting direction in perinatal health. Clinical low field (0.55T) scanners are beneficial for fetal imaging due to their reduced susceptibility-induced artefacts, increased T2* values, and wider bore (widening access for the increasingly obese pregnant population). However, the lack of standard automated image processing tools such as segmentation and reconstruction hampers wider clinical use. In this study, we introduce a semi-automatic pipeline using quantitative MRI for the fetal body at low field strength resulting in fast and detailed quantitative T2* relaxometry analysis of all major fetal body organs. Multi-echo dynamic sequences of the fetal body were acquired and reconstructed into a single high-resolution volume using deformable slice-to-volume reconstruction, generating both structural and quantitative T2* 3D volumes. A neural network trained using a semi-supervised approach was created to automatically segment these fetal body 3D volumes into ten different organs (resulting in dice values > 0.74 for 8 out of 10 organs). The T2* values revealed a strong relationship with GA in the lungs, liver, and kidney parenchyma (R\textsuperscript{2} >0.5). This pipeline was used successfully for a wide range of GAs (17-40 weeks), and is robust to motion artefacts. Low field fetal MRI can be used to perform advanced MRI analysis, and is a viable option for clinical scanning.
\end{abstract}


\section{Introduction}

Fetal magnetic resonance imaging (MRI) is becoming increasingly common, supplementing ultrasound for clinical decision-making and planning. It has a wide range of functional contrasts, a higher resolution than ultrasound, and can be used from approximately 16 weeks of gestation until birth. The renewed interest in low-field MRI (0.55T) and the increasing availability of commercial low field scanners carries significant advantages for fetal MRI: Low-field MRIs allow a wider bore (widening access to this tool for the increasingly obese pregnant population) while maintaining field homogeneity at the lower field strength, and generally do not require helium for cooling. Low field MRI is especially advantageous for fetal functional imaging (often performed using Echo-Planar-Imaging) as the reduced susceptibility-induced artefacts and longer T2* times allow for longer read-outs and hence more efficient acquisitions. It therefore provides an excellent environment for fetal body T2* mapping \cite{aviles_fast_2022}. 

T2* maps of the fetus provide an indirect measurement of blood oxygenation levels due to the differing relaxation times of deoxygenated and oxygenated hemoglobin \cite{pauling_magnetic_1936,ogawa_brain_1990}. Many fetal body organs including the lungs, kidneys, heart, liver, spleen have changing T2* values throughout gestation \cite{BAADSGAARD202390,sethi_quantification_2021}, indicating that quantitative measurements of fetal body T2* organs have the potential to be a clinically useful measurement.

However, the wider use of T2* relaxometry in the clinical setting is currently limited by significant methodological barriers such as quality of the data, fetal motion and time-consuming manual segmentations. As a consequence, it has mainly been used in research settings focused on the brain and placenta \cite{hutter_t2_2019,blazejewska_3d_2017,Schmidbauer}. Studies to date have acquired only low-resolution images of the fetal body, which do not allow for proper imaging of small organs such as the adrenal gland, and they are still susceptible to motion artifacts.
A lengthy echo time (TE) is required to acquire high-resolution multi-echo sequences. This is both impractical given the unpredictable motion and results in limited SNR on the later TEs. In structural fetal imaging, super-resolution algorithms have been used to transform several low-resolution scans into a single high-resolution volume \cite{kuklisova-murgasova_reconstruction_2012,ebner_automated_2020,uus_deformable_2020}. High-resolution reconstructions of fetal T2* maps have only been done as proof of concept in phantoms \cite{lajous_t2_2020} or in the 
 brain \cite{blazejewska_3d_2017}. Once they are acquired, the images require manual reorientation to a standard plane followed by manual segmentation in order to be clinically useful. The segmentations must also be very accurate, as inclusion of other organs or major vessels drastically changes the resulting T2* values. Overall, current state of T2* fetal body organ analysis involves a lengthy acquisition process followed by tedious manual image processing steps, resulting in a barrier to wider adoption of the technique in the clinical setting.

Here, we present an automated pipeline for quantitative mean T2* fetal body organs at low field MRI, resulting in normative growth curves from 17-40 weeks. We use a low-resolution dynamic T2* acquisition framework, and then use a novel multi-channel deformable slice-to-volume reconstruction (dSVR) to generate a high-resolution 3D volume of the fetal body and its corresponding T2* map in the standard plane \cite{uus_deformable_2020,UUS2022102484,10.1007/978-3-030-60334-2_22}, followed by automatic segmentation of the fetal body organs across a wide range of gestational ages (GA; 17-40 weeks). We generate normative T2* growth curves of ten fetal body organs at low field MRI, which has not been done at higher field strengths in such detail. This pipeline (available: https://github.com/SVRTK/Fetal-T2star-Recon) will pave the way for advanced T2* fetal body mapping to become more prevalent in both research and the clinic, potentially allowing further insights into prenatal development, and better screening and diagnostic capacities.



\section{Methods}

Fetal MRI was acquired as part of an ethically approved study (MEERKAT, REC 21/LO/0742, Dulwich Ethics Committee, 08/12/2021
) performed between May 2022 and February 2023 at St Thomas’ Hospital in London, UK
. Participants for this study were recruited prospectively, with inclusion criteria of a singleton pregnancy, maternal age over 18 years. Exclusion criteria were multiple pregnancies, maternal age $<$18 years, lack of ability to consent, weight $>$200kg, and contraindications for MRI such as metal implants. 

\subsection{Image Acquisition}
The subjects were scanned on a clinical 0.55T scanner (MAGNETOM Free.Max, Siemens Healthcare, Germany) using a 6-element blanket coil and a 9-element spine coil in the supine position. A multi-echo whole-uterus dynamic (time-resolved) gradient echo sequence was acquired in the maternal coronal orientation. The sequence parameters are as follows: Field of View (FOV): 400x400mm; resolution: 3.125mmx3.125mmx3mm; TE: [46, 120, 194] ms; TR: 10,420-18,400ms; Number of slices: 28-85 Number of dynamics: 15-30; GRAPPA: 2; flip angle 90°. The acquisition time of the dynamic T2* relaxometry scan with 20 dynamics and 3 echoes was between 4-6 minutes. 

\subsection{Image Processing}
The acquired images were reviewed to remove dynamics with excess motion. The remaining images were first denoised using MRTRIX3’s dwidenoise tool \cite{TOURNIER2019116137,CORDEROGRANDE2019391}. Next, an in-house Python script was used to generate the T2* maps for each dynamic using mono-exponential decay fitting for each voxel \cite{hutter_multi-modal_2019}. 

\subsubsection{Reconstructions} The image generated from the second echo was determined to have the best contract for the fetal body organs (an example of the echos can be seen in the Supplementary Material). This second echo image for each dynamic was then used to create a high-resolution 3D volume with isotropic resolution in a standard atlas space using deformable slice-to-volume (dSVR) registration, with the following non-default parameters: no intensity matching, no robust statistics, resolution=1.2mm, cp=[12 5], lastIter=0.015   \cite{uus_deformable_2020,UUS2022102484}. The resulting transformations are then used to create a high resolution 3D T2* map \cite{10.1007/978-3-030-60334-2_22}. 

\subsubsection{Fetal Lung Reconstruction Validation}In order to validate that the 3D volume reconstructions do not alter the obtained mean T2* values, mean T2* values from 10 of the acquired dynamics were compared with mean lung T2* value of the corresponding 3D volume. The values were considered to be equal if the mean T2* determined from the 3D volume was within two standard deviations of the mean T2* values sampled from the 10 dynamic scans.

\begin{figure}[!h]
    \centering
 \includegraphics[width=\textwidth,clip]{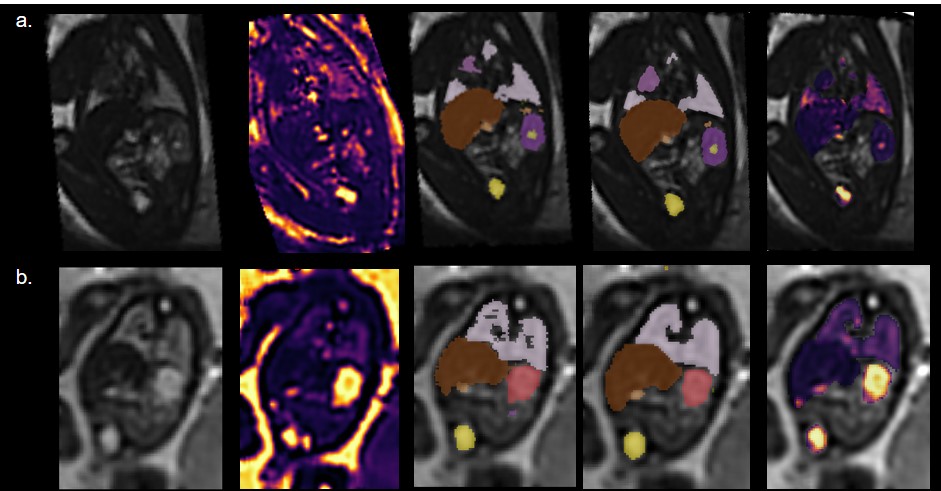}
\caption{From left to right: Example fetal deformable Slice-to-Volume Reconstructions (dSVR) of the second echo; the T2* map dSVR; the corresponding manual segmentations; label maps generated from the two-channel network; organ T2* maps overlaid on the second echo 3D volume reconstruction. Top row: 35.71 week, bottom row: 23.14 weeks} \label{segmentation}
\end{figure}

\subsubsection{Segmentation} Initial segmentations of the second echo 3D volumes were generated with an existing in-house U-Net using the MONAI framework \cite{monai_consortium_monai_2020} that had been trained on dSVR reconstructions from T2-weighted (HASTE) images of the fetal body acquired at 1.5T and 3T and corresponding manual segmentations for the lungs, thymus, gall bladder, kidney pelvis, kidney parenchyma, spleen, adrenal gland, stomach, bladder, and liver (See Figure \ref{segmentation}). Seven of the initially generated label maps were corrected in detail to create the ground truth.

A 3D nnUNet \cite{isensee_nnu-net_2021} was trained on a Tesla V100 (5 folds, 250 epochs/fold, default nnUnet data augmentation, batch size: 2, kernel size: [3,3,3], Adam optimizer, loss function: cross-entropy and Dice, 23 training cases, 4 validation cases, 7 testing cases) using these initial label maps as well as the second echo and the T2* 3D volumes. The final label maps were generated using a semi-automated refinement process, where at the end of each training cycle the cases with the best label maps were chosen, minor corrections were made by fetal anatomy specialists (Intra and inter-observer variability had previously been confirmed \cite{story_foetal_2020} before retraining. Two networks were trained, one with only the second echo 3D volumes as input (one channel), and one with both the second echo and the T2* map 3D volumes (two channel). Dice Similarity Coefficients (DSC) were calculated for both the one- and a two-channel networks and a two-sided t-test was performed.

\subsubsection{Growth Curves} Organ-specific volume and T2* growth curves were created using control cases from the generated label maps and the high-resolution T2* volumes, excluding voxels where the T2* fitting failed. A linear regression analysis was performed in order to determine the relationship between the T2* values and GA.

\begin{figure}[!h]
    \centering
    \includegraphics[width=\textwidth,clip]{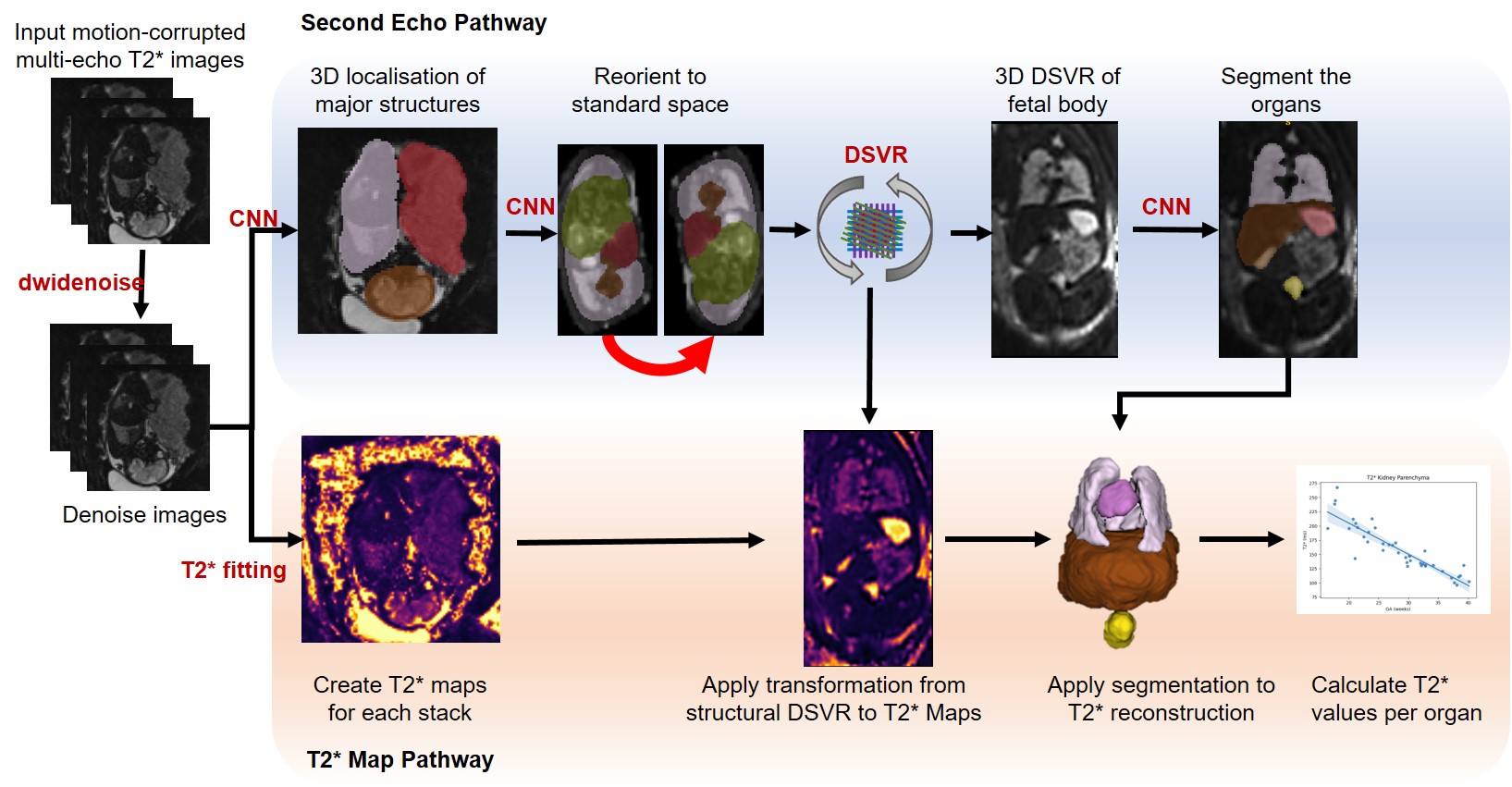}
\caption{Proposed pipeline for quantitative T2* fetal body MRI and segmentation at low field} \label{pipeline}
\end{figure}

The complete fetal body T2* 3D reconstruction and segmentation processing pipeline can be seen in Figure \ref{pipeline}.

\section{Results}

\subsection{Fetal Body T2* Reconstruction}
41 subjects had a multi-echo T2* dynamic scan (mean GA: 28.46$\pm$6.78 weeks; mean body mass index: 29.0$\pm$6.26). Dynamics with motion artifact were removed (mean included: 13.6). All cases underwent 3D volume reconstruction. Both control and pathological cases were used during the iterative training process for the segmentation network. Nine cases were excluded from the creation of growth curves due to a pathology impacting the fetal body. A further two cases were excluded after reconstruction due to excessive motion. 

\subsubsection{Fetal Lung Reconstruction Validation}
\begin{figure}[h]
    \centering
    \includegraphics[scale=0.4]{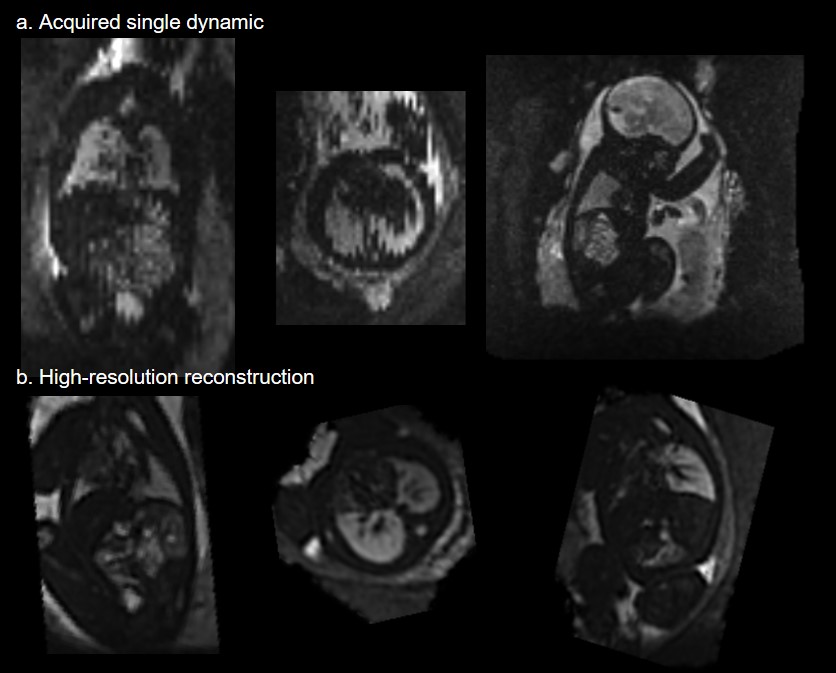}
\caption{Top row: Example of a single, motion-corrupted dynamic; Bottom Row: The reconstructed 3D volume of the same case} \label{low vs high}
\end{figure}

In the lung reconstruction validation experiment, all cases fell within the required range, and therefore are considered to be equal (Table \ref{tab2}). See Figure \ref{low vs high} for an example of the individual dynamics and the final 3D volume reconstruction.

\subsection{Fetal Body T2* Segmentation}
The DSC scores of the one- and two-channel networks (Table \ref{tab1}) showed no significant difference. The two-channel network was used to generate the T2* values in the growth curves.

\begin{table}
\caption{Mean Dice Similarity Coefficients (DSC) for each label. One Channel: Trained using the second echo 3D volume; Two Channel: Trained using both the second echo and the T2* 3D volume. No significant difference was found between the two networks.}\label{tab1}
\centering
\begin{tabular}{|l|c|c|c|}
\hline

Label & One Channel & Two Channel  & p-value\\
 & DSC mean &  DSC mean & \\
\hline
Lungs & 0.893 & 0.892 & 0.97 \\
Liver & 0.910 & 0.910 & 0.99 \\
Stomach & 0.901 & 0.900 & 0.96 \\
Spleen & 0.7934 & 0.789 & 0.95 \\
Kidney pelvis & 0.772 & 0.774 & 0.96 \\
Kidney Parenchyma & 0.847 & 0.847 & 1.00 \\
Bladder & 0.918 & 0.917 & 0.97 \\
Thymus & 0.411 & 0.416 & 0.98 \\
Gallbladder & 0.744 & 0.747 & 0.98 \\
Adrenal Glands & 0.449 & 0.453 & 0.97 \\

\hline
\end{tabular}
\end{table}

\begin{table}
\caption{Reconstruction robustness analysis: All mean lung T2* values were determined from 10 of the dynamics. All values from the reconstructed values fall within the 2$\sigma$ range, and are therefore considered to be equal.}\label{tab2}
\centering
\begin{tabular}{|l|c|c|c|}
\hline
Case   & Mean T2* & T2* 2$\sigma$ range & T2* Reconstruction\\

\hline
1  & 194.05 & 188.9 - 199.2 & 189.07\\
2  & 299.49
 & 271.9 - 327.1 & 312.25\\
3 & 225.93 & 207.9 - 244.0 & 236.09 \\
\hline
\end{tabular}
\end{table}

\subsection{Growth Curves}

\begin{figure}[]
    \centering
    \includegraphics[width=\textwidth,clip]{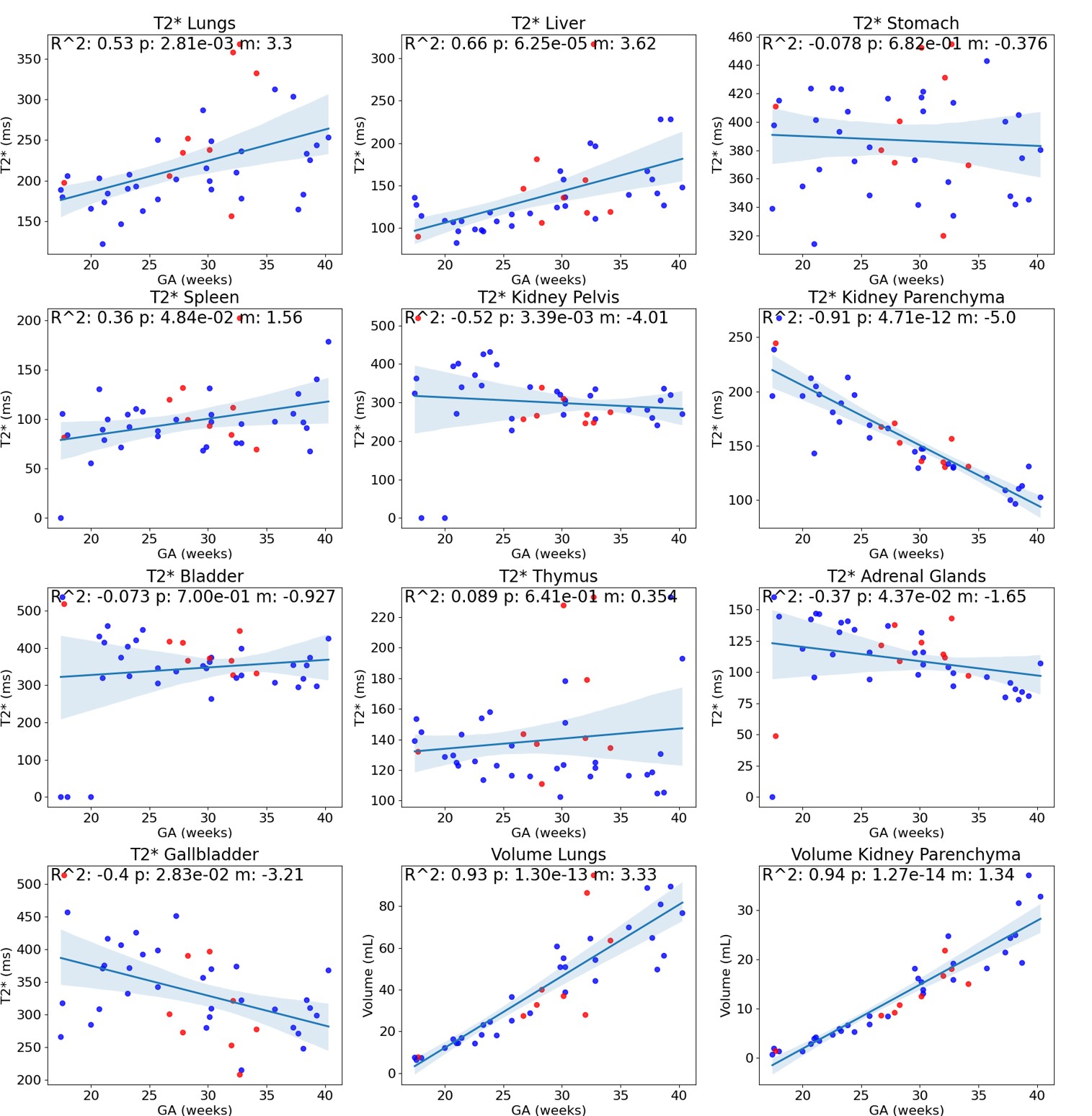}
\caption{T2* growth curves for the lungs (R\textsuperscript{2}: 0.53), liver (R\textsuperscript{2}: 0.66), stomach (R\textsuperscript{2}: -0.08), spleen (R\textsuperscript{2}: 0.36), kidney pelvis (R\textsuperscript{2}: -0.52), kidney parenchyma (R\textsuperscript{2}: -0.9), bladder (R\textsuperscript{2}: -0.07), thymus (R\textsuperscript{2}: 0.09), adrenal glands (R\textsuperscript{2}: -0.37), and gallbladder (R\textsuperscript{2}: -0.40). Two of the graphs in the bottom row (middle, right) display the lung and kidney parenchyma volumes. All organ volumes demonstrated a strong relationship with GA. Blue: Fetuses with normal organs; Red: Fetuses with a pathology potentially impacting body organs.} \label{t2s_growth}
\end{figure}

The obtained T2* growth curves (Figure \ref{t2s_growth}) show a significant increase over gestation in the lungs and liver (R\textsuperscript{2}>0.5) and a significant decrease in the kidney parenchyma and kidney pelvis (R\textsuperscript{2}>0.5). A slight relationship with GA was also found in the spleen, adrenal glands, and gallbladder. No relationship with GA was found in the stomach, bladder, and thymus. The volumetric growth curves for all organs can be found in the Supplementary Material. 

\section{Discussion and Conclusions}

The data (including images, reconstructions, and segmentations) will be made available from the corresponding author upon reasonable academic request (
). 

 The proposed pipeline overcomes barriers currently hindering wider clinical adaption such as burdensome manual image reorientation and segmentation. It requires users only to acquire the images and review the input scans for motion before starting the pipeline. This automation would allow the pipeline to move advanced fetal image analysis outside of specialist centers and into a more standard workflow. The pipeline worked from 18-40 GA, with only 2 cases (which were <20 weeks) discarded due to motion, and was able to confirm literature trends for all organs except the liver and lungs \cite{sethi_quantification_2021,BAADSGAARD202390}, where the opposite trend was observed. This may be due to the different GA ranges included in the literature. The increasing mean T2* values in the lungs may be due to the increasing vascularity, thereby increasing the amount of fetal haemoglobin present. The long T2* values achieved with the low field MRI allows for excellent contrast in the fetal body, which assists in the reconstruction. This pipeline will provide insight into the development of fetal body organs not yet explored, such as the adrenal glands. A more comprehensive study with more cases at every GA is needed to further validate these normative curves.

While structural T2-weighted images are more suited for volumetry, the fact that all organs follow the expected growths trend further validates the segmentations. It also indicates that for some of the smaller organs (kidney pelvis, adrenal glands), our network has difficulties in the segmentation step for younger fetuses. Two organs (thymus, adrenal glands) had poor DSC values (below 0.45), indicating that further work on the segmentation network for these difficult organs is required. The thymus has very poor contrast and is difficult to delineate even manually. The network often identified heart tissue as thymus, skewing the T2* values calculated. The adrenal gland is a very small organ, which makes it difficult to segment. The poor DSC for the adrenal gland does not necessarily translate to incorrect average T2* values, as the DSC is a volumetric metric. However, improved DSC scores would allow for more confidence in the calculated T2* values.

The proposed multi-organ pipeline can be successfully run across a wide range of GAs, and requires minimal user interaction. The combination of low field fetal MRI, quantitative imaging, and comprehensive image analysis pipelines could potentially make a substantial impact in our understanding of the development of the fetal body throughout gestation, as well as possibly provide clinical prenatal biomarkers. 

\section{Acknowledgments}
The authors thank all the participating families as well as the midwives and radiographers involved in this study. This work was supported by the the NIH (Human Placenta Project—grant 1U01HD087202‐01), Wellcome Trust Sir Henry Wellcome Fellowship (201374/Z/16/Z and /B), UKRI FLF (MR/T018119/1), the NIHR Clinical Research Facility (CRF) at Guy's and St Thomas' and by core funding from the Wellcome/EPSRC Centre for Medical Engineering [WT203148/ Z/16/Z]. For the purpose of Open Access, the Author has applied a CC BY public copyright license to any Author Accepted Manuscript version arising from this submission. The views expressed are those of the authors and not necessarily those of the NHS or the NIHR.


\bibliographystyle{splncs04}
\bibliography{MICCAI.bib}
\end{document}